\documentclass[pra,showpacs,amsmath,amssymb,singlecolumn,superscriptaddress,twocolumn]{revtex4}
\usepackage{bm,multirow}
\usepackage[usenames]{color}
\newcommand \beq{\begin{eqnarray}}
\newcommand \eeq{\end{eqnarray}}
\def\simge{\mathrel{%
       \rlap{\raise 0.511ex \hbox{$>$}}{\lower 0.511ex \hbox{$\sim$}}}}
\def\simle{\mathrel{
       \rlap{\raise 0.511ex \hbox{$<$}}{\lower 0.511ex \hbox{$\sim$}}}}
\usepackage{graphics}
\usepackage[dvips]{graphicx}
\usepackage{graphicx}
\newcommand{\qq}{\mathbf{q}}
\newcommand{\pp}{\mathbf{p}}

\newcommand{\rr}{\mathbf{r}}

\newcommand{\be}{\begin{equation}}
\newcommand{\ee}{\end{equation}}
\newcommand{\bea}{\begin{eqnarray}}
\newcommand{\eea}{\end{eqnarray}}
\newcommand{\ba}{\begin{align}}
\newcommand{\ea}{\end{align}}

\begin{document}
\title{Clock shifts in a Fermi gas interacting with a minority component: a soluble model}
\author{G.\ M.\  Bruun}
\affiliation{NORDITA, Roslagstullsbacken 23, SE-10691 Stockholm, Sweden}
\affiliation{Mathematical Physics, Lund Institute of Technology, P.\ O.\ Box 118, SE-22100 Lund, Sweden}
\author{C.\ J.\ Pethick}
\affiliation{NORDITA, Roslagstullsbacken 23, SE-10691 Stockholm, Sweden}
\affiliation{The Niels Bohr International Academy, The Niels Bohr Institute, Blegdamsvej 17, DK-2100 Copenhagen \O, Denmark}
\author{Zhenhua Yu}
\affiliation{The Niels Bohr International Academy, The Niels Bohr Institute, Blegdamsvej 17, DK-2100 Copenhagen \O, Denmark}

\date{\today}

\begin{abstract}
We consider the absorption spectrum of a Fermi gas mixed with a minority species when majority fermions are transferred to another internal state by  an external probe.
 In the limit when the minority species is much more massive than the majority one, we show that the minority species may be treated as static impurities
  and the problem can be solved in closed form.   The analytical results bring out the importance of vertex corrections, which change qualitatively the nature of the absorption spectrum.
  It is demonstrated that large line shifts are not associated with resonant interactions in general.
  We also show that  the commonly used ladder approximation fails when the majority component is degenerate for large mass ratios between the minority and majority species and that bubble diagrams, which correspond to the creation of many particle--hole pairs, must be taken into account.
We carry out detailed numerical calculations, which confirm the analytical insights and
we point out the connection to shadowing phenomena in nuclear physics.

\pacs{03.75.Ss, 05.30.Fk, 67.85.Jk }

\end{abstract}

\maketitle

\section{Introduction}
The spectroscopy of spin excitations in atomic systems is important  for basic science as well as being
 technologically relevant to atomic clocks.  The subject has a long history, going back to studies of spin-exchange optical pumping \cite{balling} and of line shifts in hydrogen masers \cite{koelman}.  In recent years it has acquired renewed interest following experiments on ultracold atomic gases that have played an important role in probing effects of interatomic interactions in these systems
 \cite{gupta, mit_two, chin_rf,  shin}.  In a typical experiment, one
induces transitions of atoms from one hyperfine state of the ground state manifold, denoted by 1, to a second hyperfine state, 2, in the presence of atoms in a third state, 3.
Particular interest has focused on situations where the interatomic interactions are strong, for example for $^6$Li for which scattering lengths have magnitudes $\sim 10^3 a_0$ for a large range of magnetic fields.

The quantity measured in experiment is basically a two particle correlation function that is difficult to calculate when interactions are strong.    Many effects have to be considered, including particle self energies, vertex corrections, pairing, and the inhomogeneity of the atomic cloud \cite{kinnunen,yu,perali,pieri,levin,baymzwierlein,zwerger,bruun1,bruun2,mueller}.
In this paper we consider a simple model where the mass of the bystander atom, 3, is much larger that
of states 1 and 2.  This allows us to include self energy and vertex corrections to all orders in a conserving approximation which becomes exact when the system is highly polarized, in the sense that the density of the bystander atoms is much smaller than the density of the 1 atoms.  Throughout most of the paper we shall neglect the interaction between 1- and 2-atoms which does not give rise to shifts in the absence of interactions with 3-atoms. For densities of 3-atoms low enough that they are nondegenerate, the statistics of these atoms plays no role, so our calculations apply to both bosons and fermions.

For the case of nondegenerate majority atoms, our formalism enables us to derive in a straightforward way analytical results obtained previously  \cite{balling,koelman}.  The calculations bring out the important role of the processes analogous to those considered by Aslamazov and Larkin \cite{AL} in studies of fluctuation contributions to response function close to the transition temperature in superconducting metals.

We find that vertex corrections can qualitatively change the clock shift compared with the prediction without vertex corrections. We also show that large
line shifts are not associated with resonant interactions. For instance,  when one interaction, e.g.\ 1-3, is
on resonance, the clock shift has the same magnitude {\it but the opposite sign} compared with its value
when the 1-3 interaction is zero. Our analytical results are confirmed  by numerical calculations.
Another conclusion of the work is that for the case of massive bystander atoms, the common approximation of including only ladder diagrams is inadequate, since  particle-hole correlations must be considered on the same footing at particle-particle and hole-hole correlations.  We also discuss the relationship of the physics of the clock shift problem to the phenomenon of ``shadowing'' in nuclear physics, the fact that, e.g., the total cross section for scattering of a pion from the deuteron is not equal to the sum of the cross section for scattering from a proton  and that for scattering from a neutron \cite{glauber}.

This paper is organized as follows.  In Sec.\ II we describe the basic formalism for calculating the transition rate, and in Sec.\ III we describe the calculation of the line shape under the assumption that the massive atoms may be treated as static impurities.  After deriving analytical result we present results of numerical calculations.  Section IV is devoted to showing from diagrammatic perturbation theory that for a mobile minority species with a large mass, the problem reduces to that of scattering from static impurities.  There we also consider the relationship of our calculations to the X-ray edge problem and the phenomenon of  ``shadowing'' in nuclear physics.  Finally, Sec.\ V contains concluding remarks.

\section{Transition rate}
We consider a gas of fermions in an internal state $1$ with density $n_1$ and mass $m$  which interacts with
a gas of fermions or bosons of mass $m_3$ and density $n_3$ which is assumed to be much smaller than  $n_1$.      The gas is subjected to a homogeneous
 probe that flips the fermions from state $1$ to state $2$ at a rate which within
 linear response theory is proportional to
 \be
 \sum_{i,f}(P_i-P_f)\left|\int d^3r\langle f| \psi_2^\dagger (\rr) \psi_1(\rr)    |i\rangle\right|^2\delta(\omega-E_f+E_i),
 \ee
 where initial states are denoted by $i$ and final ones by $f$, and their energies by $E_i$ and $E_f$.  The frequency of the applied field is $\omega$. (We put $\hbar$ and the Boltzmann constant equal to unity throughout.) The probability of occupation of the initial (final) state is denoted by $P_i$ ($P_f$).
   The operator $\psi_\sigma^\dagger(\rr)$ creates  a fermion in state $\sigma$ at position $\rr$.
In terms of correlation functions, the rate is proportional to
 \begin{equation}
{\rm Im} {\mathcal{D}}(\omega)\propto \int d\mathbf{r}d\mathbf{r}'{\rm Im}{\mathcal{D}}({\mathbf{r}},{\mathbf{r}}',\omega)
\end{equation}
where ${\mathcal{D}}({\mathbf{r}},{\mathbf{r}}',\omega)$ is the Fourier transform of the quantity \newline
$-i\theta(t-t')\langle[\psi_2^\dagger({\mathbf{r}},t)\psi_1({\mathbf{r}},t),\psi_1^\dagger ({\mathbf{r}}',t')\psi_2({\mathbf{r}}',t')]\rangle$ which may be regarded as the correlation function for the pseudospin operator that describes atoms in the states 1 and 2.

\section{Static impurities}
\label{impurity}

\begin{figure}
\begin{center}
\includegraphics[width=\columnwidth]{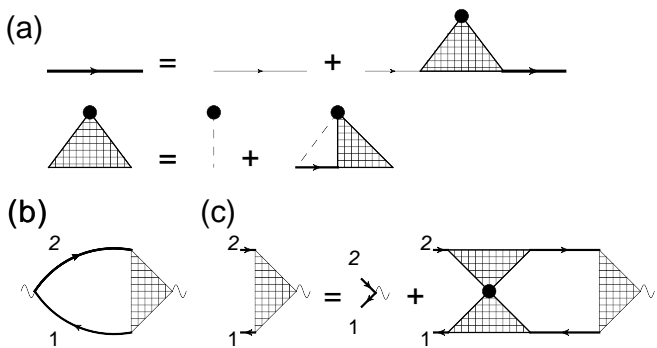}
\caption{(a) The propagator $G$ for the fermions scattering on impurities. (b) The
correlation function ${\mathcal{D}}(\omega)$. (c) The vertex function.
Thick solid lines indicate $G$, thin solid lines $G_0$, and dashed lines scattering on a impurity marked
by $\bullet$.}
\label{Feynman}
\end{center}
\end{figure}

In this section, we consider the case when $m_3\gg m$ so that the
3-atoms may be treated as static impurities, as we shall demonstrate
in Sec.\ IV.
 We shall work at nonzero temperature $T$, in which case the frequencies are to be regarded initially as Matsubara frequencies,
 odd multiples of $\pi T$ for fermions  and even multiples for bosons.  The real time correlation function is then
obtained in the standard way by analytically continuing from the imaginary time domain.
The fermion propagators in the presence of the impurities are
\be
G_\sigma(p,z)^{-1}=G_\sigma^0(p,z)^{-1}-\Sigma_\sigma(z),
\label{G}
\ee
where
\be
G_\sigma^0 (p,z)^{-1}= z-p^2/2m -\epsilon_\sigma +\mu_\sigma
\label{G0}
\ee
is the bare propagator and $\Sigma_\sigma(z)$ the self energy. Here $\epsilon_\sigma$ is the energy of a noninteracting $\sigma$-fermion ($\sigma=1,2$)
at rest and $\mu_\sigma$  the chemical potential.   In Eq.\ (\ref{G}) it is understood that the propagator is averaged over a random distribution of impurities \cite{AGD}, but we shall not indicate this explicitly in the notation.  A similar remark applies to  the correlation function ${\mathcal{D}}(\omega)$.

To lowest order in $n_3$, the self energy has the form \be
\Sigma_\sigma(z) =n_3 {\mathcal{T}}_{\sigma}(z). \label{sigma} \ee
Here ${\mathcal{T}}_{\sigma}(z)$ is the $T$-matrix for scattering of
a $\sigma$-fermion on an impurity, which is given in a matrix
notation by \be {\mathcal{T}}_{\sigma}(z)=V_\sigma +V_\sigma
G_\sigma(z){\mathcal{T}}_{\sigma}(z), \label{lippmann} \ee where the
momentum sums are implicit. We have assumed that the range of the
interaction $V_\sigma$ between the impurities and the fermions is
much shorter than the lesser of the typical interparticle distance
and the thermal de Broglie wavelength, $(2\pi/mT)^{1/2}$. In this
case, for the momenta of interest,
 the scattering  amplitude  depends only on
the energy.  Equations (\ref{G})-(\ref{sigma}) are shown diagrammatically in Fig.\ \ref{Feynman}(a).

The correlation  function ${\mathcal{D}}(\omega)$ is shown
diagrammatically in Fig. \ref {Feynman}(b), and the vertex function
in Fig. \ref{Feynman}(c). The importance of vertex corrections may
be illustrated by considering the case when the interaction between
an impurity and a fermion is the same for the two fermion species.
The Hamiltonian is then SU$(2)$ symmetric with respect to rotations
between the states $1$ and  $2$ and the correlation function
${\mathcal{D}}(\omega)$ is unaffected by interactions \cite{yu}.
However, the self energy corrections to the single-particle
propagator are nonzero and they must therefore be canceled by the
vertex corrections.  To recover the SU$(2)$ invariance  and to
satisfy conservation laws, it is necessary when  calculating the
correlation function to use as the vertex in the particle-hole
channel the quantity $\delta \Sigma/\delta G$ ($\Sigma$ and $G$ are
matrices in 1-2 space)~\cite{baym, yu}.
 The structure of the vertex corrections  is shown in Fig.\ \ref{Feynman}(c).
We now perform such a conserving calculation of
${\mathcal{D}}(\omega)$ which  takes the effects of the impurities into account exactly to lowest order in $n_3$.

 The vertex corrections correspond to processes in which a 2-particle and a 1-hole  scatter
  from the same impurity.   The resulting effective interaction
  [the ``bow-tie" part of the diagram in Fig.\ \ref{Feynman}(c)] is given by
\be
V_{\rm eff}= n_3{\cal T}_1(  i\omega_\nu     ){\cal T}_2(i\omega_\nu+i\omega_\gamma) ,
\label{veff}
 \ee
  since for a static impurity the energy transfer between particles and the impurity is zero.
Because the scattering on an impurity is  independent of momentum, the inclusion of the vertex corrections simplifies significantly, and
 the diagrams for the  correlation function may be summed. The result is
\begin{widetext}
 \begin{equation}
{\mathcal{D}}(i\omega_\gamma)=T\sum_{\omega_\nu}\frac{(2\pi)^{-3}\int d^3p \,G_1(p,i\omega_\nu)G_2(p,i\omega_\nu+i\omega_\gamma)}
{1-n_3{\mathcal{T}}_1(i\omega_\nu){\mathcal{T}}_2(i\omega_\nu+i\omega_\gamma)(2\pi)^{-3}\int d^3p \,G_1(p,i \omega_\nu)G_2(p,i\omega_\nu+i\omega_\gamma)}
\label{Startchi}
\end{equation}
\end{widetext}
with  $\omega_\nu$ being a fermion Matsubara frequency and $\omega_\gamma$ a boson one.
The momentum integral yields
 \begin{gather}
M(z_1,z_2)=\int\frac{d^3p}{(2\pi)^3}G_1(p,z_1)G_2(p,z_2)\nonumber\\
=i\pi \frac{d_2(z_2){\rm sgn}({\rm Im} z_2)-d_1(z_1){\rm sgn}({\rm Im} z_1)}{z_2-z_1+\mu_2-\mu_1-\Delta+\Sigma_1(z_1)-\Sigma_2(z_2)}
\end{gather}
with $d_\sigma(z)=m^{3/2}\sqrt{z+\mu_\sigma-\epsilon_\sigma-\Sigma_\sigma(z)}/\sqrt 2\pi^2$ and $\Delta=\epsilon_2-\epsilon_1$
 the hyperfine splitting between the two fermionic states.  For $\Sigma=0$, $d$ is the free particle density of  states.
We evaluate the sum over Matsubara frequencies in (\ref{Startchi}) by converting it to a contour integration in the usual way by multiplying the integrand by the Fermi function
$f(z_1)= [\exp(\beta z_1)+1]^{-1}$
and choosing a contour that encircles the poles of the Fermi function.   The integration contour may be deformed  to lie above and below the cuts of the functions $d_1$ and $d_2$, which are located at $z_1=\epsilon$ and $z_1=\epsilon-i\omega_\gamma$,
respectively, where $\epsilon$ is real.
After the analytic continuation $i\omega_\gamma\rightarrow {\tilde\omega}+i\eta$, with  the physical frequency
given by $\omega=\tilde\omega+\mu_2-\mu_1$, we obtain
 \begin{gather}
{\mathcal{D}}(\omega)=
-\int_{-\infty}^\infty \frac{d\epsilon}{2\pi i} f(\epsilon)\left[
{\mathcal{S}}(\epsilon+i\eta,\epsilon+\tilde\omega+i\eta)-\right.\nonumber \\\left.
{\mathcal{S}}(\epsilon-i\eta,\epsilon+\tilde\omega+i\eta)
+{\mathcal{S}}(\epsilon-\tilde\omega-i\eta,\epsilon+i\eta)\right.\nonumber \\\left.
-{\mathcal{S}}(\epsilon-\tilde\omega-i\eta,\epsilon-i\eta)\right],
\label{Full}
\end{gather}
where
\begin{equation}
{\mathcal{S}}(z_1,z_2)=\frac{M(z_1,z_2)}{1-n_3{\mathcal{T}}_1(z_1){\mathcal{T}}_2(z_2)M(z_1,z_2)}.
\label{S}
\end{equation}
The imaginary part of the  correlation function becomes
\begin{widetext}
 \begin{gather}
{\rm Im}{\mathcal{D}}(\omega)=\int\frac{d\epsilon}{2}(f_2-f_1)
{\rm Im}
\left[
\frac{d_2- d_1}
{\omega-\Delta+n_3[{\mathcal{T}}_1-{\mathcal{T}}_2-i\pi {\mathcal{T}}_1{\mathcal{T}}_2(d_2- d_1)]}
-\frac{d_2+d_1^*}
{\omega-\Delta+n_3[{\mathcal{T}}_1^*-{\mathcal{T}}_2-i\pi {\mathcal{T}}_1^*{\mathcal{T}}_2(d_2 +d_1^*)]}
\right].
\label{GrandFormula}
\end{gather}
\end{widetext}
In (\ref{GrandFormula}), $f_1=f(\epsilon)$,
$f_2=f(\epsilon+\tilde\omega)$ and $d_1$ and ${\mathcal{T}}_1$ are
evaluated at the energy $\epsilon+i\eta$ and $d_2$ and
${\mathcal{T}}_2$ at the energy $\epsilon+\tilde\omega+i\eta$. The
$T$-matrix, given by Eq.\ (\ref{lippmann}), has the form \be {\cal
T}_\sigma =\frac{{\cal T}_{\sigma{\rm vac}}}{1+i \pi d_\sigma{\cal
T}_{\sigma{\rm vac}}}, \ee
 where ${\cal T}_{\sigma{\rm vac}}$ is the  $T$-matrix for $\sigma$-fermions scattering at zero energy in a vacuum.
 From this it follows that  Eq.\  (\ref{GrandFormula}) reproduces the unshifted ideal gas result when the interaction
 is SU$(2)$ symmetric with identical scattering between an impurity and the two fermionic species. Note that it is crucial to use full propagators to recover this symmetry.

\subsection{A simple limit}\label{Simpel}
Equation (\ref{GrandFormula}) satisfies the conservation laws regardless of the magnitude of $n_3$. We now study the interaction effects on ${\mathcal{D}}(\omega)$ 
to the lowest order in $n_3$ and neglect all medium effects except the factor $n_3$ in front of the $T$-matrices in the denominator  in (\ref{GrandFormula}). 
The $T$-matrix is thus replaced by its value in a vacuum given by 
 \be {\cal T}_\sigma =
i\frac{e^{2i\delta_\sigma}-1}{2\pi d_0}, \ee
 where $\delta_\sigma$ is the scattering phase shift in a vacuum and $d_0=m^{3/2}\sqrt{\epsilon}/\sqrt 2\pi^2$ is the free-particle density of states. For low energies, the phase shift is given in terms of the scattering length $a_\sigma$ by $\tan \delta_\sigma=-k a_\sigma$, where $k=\sqrt{2m\epsilon}$.

  We make the variable change $\epsilon+\mu_1-\epsilon_1\rightarrow \epsilon$, so that $\epsilon$ is the kinetic energy of a 1-fermion.  The 2-fermion has kinetic energy $\epsilon +\omega-\Delta$.  In the limit of a low density of 3-atoms, the line shifts  are small. One may then
  neglect differences between $\omega$ and $\Delta$, and therefore the phase shifts of the two fermions are to be evaluated at the same kinetic energy.
 In total, keeping only the lowest order effects of $n_3$ in  (\ref{GrandFormula})  yields
\begin{equation}
{\rm Im}{\mathcal{D}}(\omega)\simeq {\rm Im}\int_0^\infty d\epsilon
\frac{-d_0\,f(\epsilon+\epsilon_1-\mu_1)}
{\omega-\Delta -  in_3\left[e^{2i(\delta_1-\delta_2)}-1\right]/2\pi d_0}.
\label{Lorentz}
\end{equation}
We have written  Eq.\ (\ref{Lorentz}) for  the case  where there are no 2-fermions present initially
so that the first  Fermi function in Eq.\ (\ref{GrandFormula}) is zero. This
corresponds to a typical experimental situation.
For equal interaction between the an impurity and the initial $1$- and final $2$-fermions, i.e.\ for $\delta_1=\delta_2$,
ones sees immediately that the interaction effects vanish in the denominator of (\ref{Lorentz}) and we recover the ideal gas result
  \begin{equation}
{\mathcal{D}}(\omega)=-\frac{n_1}{\omega-\Delta+i\eta}.
\label{IdealGas}
\end{equation}
In general,  ${\rm Im} {\mathcal{D}}(\omega)$ is the sum of Lorentzian lines, with the energy-dependent
frequency shift
 \begin{equation}
 \Delta \omega(\epsilon)=n_3\frac{\pi }{m}\frac{\sin(2\delta_{1}-2\delta_{2})}{k},
 \label{ShiftVertex}
\end{equation}
and with full width at half maximum equal to
 \begin{equation}
\Gamma(\epsilon)=n_3\frac{4\pi }{m}\frac{\sin^2(\delta_{1}-\delta_{2})}{k}.
\label{WidthVertex}
\end{equation}
Equations (\ref{ShiftVertex})-(\ref{WidthVertex}), which apply for arbitrary degree of degeneracy of the 1-atoms, have a form similar to those
derived for a classical gas in Refs.\  \cite[Eq.\ (82)]{balling} and \cite{koelman}, which studied the equation of motion for the density matrix.  There is, however, a difference, since to obtain the result in these papers one must replace the term containing the phase shifts in the denominator of Eq.\ (\ref{Lorentz}) by its thermal average.
Since $\Delta$ and $\Gamma$ are energy dependent,  the line shape given by Eq.\ (\ref{Lorentz})
will not be Lorentzian in general. The magnitude of deviations from Lorentzian behavior will depend on the variation of  $  [e^{2i(\delta_1-\delta_2)}-1]/k$,  over the distribution of the momentum $k$ of 1-atoms.

For small phase shifts, $\delta_\sigma \simeq k a_\sigma$ and  Eq.\ (\ref{ShiftVertex}) reproduces the low density expression for the shift, $\Delta\omega \simeq n_3 2\pi (a_2-a_1)/m$, the factor of two, rather than the usual factor of four for the case of fermions of equal mass,  being due to the fact that we have taken the 3-atoms to be infinitely massive.

Equations (\ref{ShiftVertex})-(\ref{WidthVertex}) clearly illustrate the importance of vertex corrections.
When these are neglected, the corresponding results are
 \begin{equation}
 \Delta \omega(\epsilon)|_{\rm no\,vertex}=n_3\frac{\pi }{m}\frac{\sin 2\delta_{1}-\sin 2\delta_{2}}{k},
 \label{ShiftNoVertex}
\end{equation}
and
 \begin{equation}
\Gamma(\epsilon)|_{\rm no\,vertex}=n_3\frac{4\pi }{m}\frac{\sin^2\delta_{1}+\sin^2\delta_{2}}{k}.
\label{WidthNoVertex}
\end{equation}

There are a number of important conclusions that may be drawn from the above results.
First, without vertex corrections the shift and damping do not display the required SU$(2)$ symmetry for $\delta_1=\delta_2$. Second, for small phase shifts the line width is proportional to $(a_1-a_2)^2$, whereas without vertex corrections the corresponding result is $a_1^2 + a_2^2$.  Thus, even in the limit of  small phase shifts it is important to include vertex corrections, which give an  interference term $-2a_1a_2$.
Third, the largest shifts are obtained for $\delta_1-\delta_2 \approx \pi/4+\nu\pi/2$, where $\nu$ is an integer.  Thus, resonant scattering is not particularly favorable for producing large shifts.  For example, take a typical experimental situation where scattering of the fermions in the initial state 1 with 3-atoms is resonant while that of final state 2 fermions is not:  the shift is then equal in magnitude but {\it of the opposite sign} compared with what it would be in the absence of 1-3 scattering, and therefore the magnitude of the shift is determined completely by the non-resonant 2-3 interaction.  Finally, large widths and large shifts do not go hand in hand, since the largest widths occur when $\delta_1-\delta_2$ is an odd multiple of  $\pi/2$.

\subsection{Numerical results}
\begin{figure}
\begin{center}
\includegraphics[width=\columnwidth]{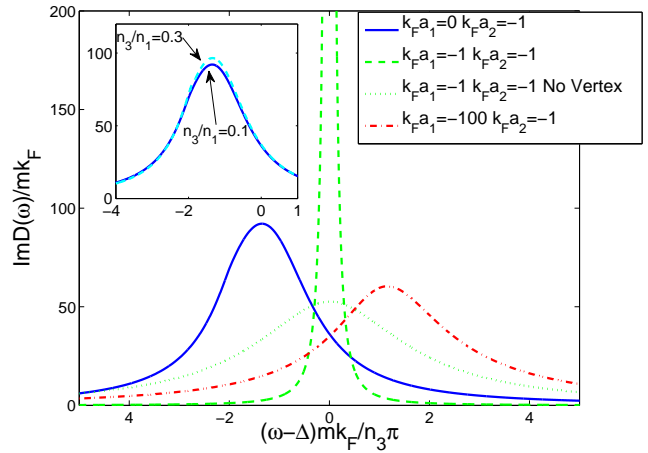}
\caption{The transition rate as a function of frequency and interaction. The inset shows the transition rate for varying impurity concentration.}
\label{SU2}
\end{center}
\end{figure}

\begin{figure}
\begin{center}
\includegraphics[width=\columnwidth]{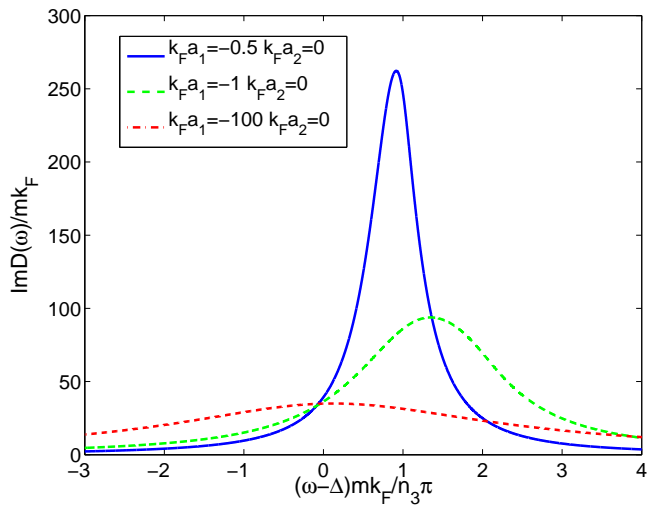}
\caption{The transition rate as a function of frequency for varying initial state interaction.}
\label{ShiftvsDamp}
\end{center}
\end{figure}

In Figs.\ \ref{SU2}-\ref{ShiftvsDamp} we present numerical results for ${\rm Im}{\mathcal{D}}(\omega)$
 obtained from Eq.\ (\ref{GrandFormula}). The propagators used in this calculation are determined fully self-consistently. 
 The frequency unit is $\pi n_3/mk_F$ with $n_1=k_F^3/6\pi^2$
 [see (\ref{ShiftVertex})-(\ref{WidthVertex})], and ${\rm Im}{\mathcal{D}}(\omega)$ plotted in units of $mk_F$.
  In Fig.\ \ref{SU2}, we show the transition rate for $T=0$
 and with an impurity density $n_3/n_1=0.1$. The scattering length for the 2-3 interaction is $a_2=-1/k_F$
 and the scattering length for the 1-3 interaction  varies from $a_1=0$ to  $a_1=-100/k_F$, which is very close to resonance.
 For $k_Fa_1=0$, the line shift and width are due solely to the self energy of the 2-atom.
 We see that when  the 1-3 interaction
 is resonant, the main effect  is to change the sign of line shift compared with the result for  zero 1-3 interaction. This confirms
  the discussion in Sec.\ \ref{Simpel}. When $k_Fa_1=k_Fa_2=-1$, the unshifted
 ideal gas result is recovered, as it should be, and the small remaining width of the calculated signal is entirely due to a
  small imaginary part we have added explicitly to the frequency to facilitate the numerical calculations.  For comparison, we also plot the result obtained for
 $k_Fa_1=k_Fa_2=-1$ when vertex corrections are not included. We see that, although the predicted line shift is small, the width is large
 and one does not recover the unshifted narrow line when vertex corrections are ignored.

 In the inset, we compare the line shape for $n_3/n_1=0.1$
 and $n_3/n_1=0.3$, keeping  $k_Fa_1=0$ and $k_Fa_2=-1$ fixed.  To ease comparison of the results for the two different impurity concentrations, we have
   multiplied ${\rm Im}{\mathcal{D}}(\omega)$  by $n_3$. The two curves largely overlap which
   illustrates that the line shift and width  essentially scale with $n_3$ in agreement with (\ref{ShiftVertex})-(\ref{WidthVertex}). Note that 
higher order medium effects coming from the self-consistent determination of the propagators give rise to the  slight difference between the results for the two 
impurity concentrations.

In Fig.\ \ref{ShiftvsDamp}, we plot the transition rate as a function of  $a_1$ keeping $a_2=0$. The line shift is large  whereas the width is
small for $k_Fa_1=-1$. This is in agreement with the conclusions reached in Sec.\ \ref{Simpel} from (\ref{ShiftVertex})-(\ref{WidthVertex}) since $k_Fa_1=-1$
 corresponds to a phase shift of $\delta_1=\pi/4$. Likewise when the scattering is close to resonance with $k_Fa_1=-100$ corresponding
to $\delta_1\approx \pi/2$, the line shift is small whereas the width is large. Again, this agrees with the discussion in Sec.\ \ref{Simpel}

\section{Mobile, massive impurities}

In this section we show that in the limit $m/m_3\rightarrow 0$, the results we have employed in Section \ref{impurity} for the self-energy and vertex corrections  may be derived from diagrammatic many-body theory for particles of finite mass.  An important conclusion of this section is that the ladder approximation, which is commonly employed in treating strongly interacting systems is inadequate to describe systems with $m_3\gg m$ when the 1-fermions are degenerate. We begin by showing this for low-order contributions in perturbation theory, and then generalize the considerations to arbitrary order.  For definiteness, we shall assume that the 3-atoms are fermions, but the calculations may easily be generalized to the case of bosons, the only difference being that the distribution function for 3-atoms must be taken to be the Bose distribution.  To lowest order in the density of 3-atoms, the results are independent of the statistics of the 3-atoms.

\subsection{Second order}
\begin{figure}
\begin{center}
\includegraphics[width=\columnwidth]{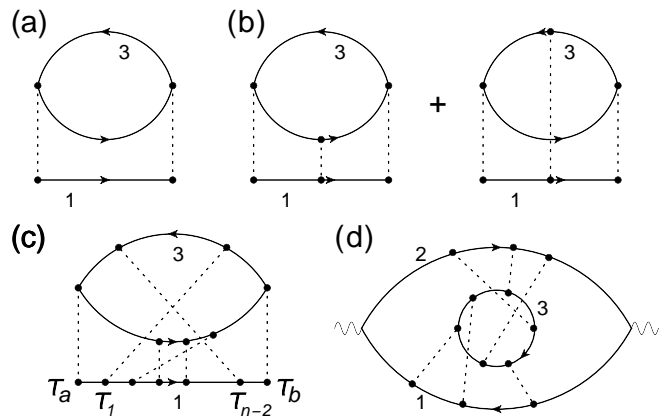}
\caption{Diagrams for the self energy of a 1-fermion. (a) Second order contributions. (b) Third-order contributions.  (c) A general term.
(d) A diagram with a general vertex correction. Solid lines are fermion propagators, dashed lines interactions with $\bullet$ indicating the vertices.}
\label{Sigmadiagrams}
\end{center}
\end{figure}

In the Hartree approximation the self energy and vertex corrections
are given by Eqs.\ (\ref{sigma}) and (\ref{veff}) when the
$T$-matrix is evaluated in the Born approximation, so the first term
we shall consider in detail is the second-order term, Fig.\
\ref{Sigmadiagrams}(a).
For $m/m_3\rightarrow0$ it is given by
\begin{gather}
\Sigma_1^{(2)}(p,\omega)=
 \int\frac{d^3q}{(2\pi)^3}\frac{d^3p'}{(2\pi)^3}
 |V_1(q)|^2\times\nonumber\\
 \frac{ f^3_{\pp'}(1-f^3_{\pp'+\qq}) (1-f^1_{\pp-\qq}) + (1-   f^3_{\pp'})f^3_{\pp'+\qq} f^1_{\pp-\qq}    }{\omega -(\pp-\qq)^2/2m},
\end{gather}
where $V_\sigma(\qq)$ is the bare interaction between a 3-atom and a $\sigma$-fermion and $f^i_\pp=f(p^2/2m_i+\epsilon_i-\mu_i)$.  When the 3-atoms are nondegenerate the $1-f^3$ factors may be replaced by unity,
and one then finds
\begin{gather}
\Sigma_1^{(2)}
=
{\int\frac{d^3q}{(2\pi)^3}\frac{d^3p'}{(2\pi)^3}}
|V_1(q)|^2\frac{ f^3_{\pp'} (1-f^1_{\pp-\qq}) + f^3_{\pp'+\qq} f^1_{\pp-\qq}    }{\omega -(\pp-\qq)^2/2m} \nonumber\\
=n_3
{\int\frac{d^3q}{(2\pi)^3}}
 |V_1(q)|^2\frac{  (1-f^1_{\pp-\qq}) +  f^1_{\pp-\qq}    }{\omega -(\pp-\qq)^2/2m}  =
n_3{\cal T}^{(2)}_{\rm vac}
\end{gather}
where
\begin{gather}
{\cal T}^{(2)}_{\rm vac}(p,\omega)={\cal T}^{(2)}_{\rm lad}(p,\omega) +{\cal T}^{(2)}_{\rm bub }(p,\omega)\nonumber\\
=\int\frac{d^3q}{(2\pi)^3} |V_1(q)|^2\frac{1   }{\omega -(\pp-\qq)^2/(2m)}
\end{gather}
is the $T$-matrix  in a vacuum (calculated to second order in
$V_1$). Here \be
 {\cal T}^{(2)}_{\rm lad}(p, \omega)  =      \int\frac{d^3q}{(2\pi)^3} |V_1(q)|^2\frac{  1-f^1_{\pp-\qq}   }{\omega -(\pp-\qq)^2/(2m)}
\ee is the contribution to the $T$-matrix from ladder diagrams
(particle-particle scattering and hole-hole scattering) and \be
{\cal T}^{(2)}_{1,\rm bub}(p,\omega)= \int\frac{d^3q}{(2\pi)^3}
|V_1(q)|^2\frac{   f^1_{\pp-\qq}    }{\omega -(\pp-\qq)^2/(2m)} \ee
is the contribution from particle-hole scattering.  This calculation
leads to two important conclusions.  First, the presence of the
degenerate 1-fermions affects scattering in the particle-particle
and hole-hole channel and also scattering in the particle-hole
channel, but the effects of occupancy of intermediate states cancel
in the total, which to second order is given in terms of the
$T$-matrix {\it in vacuo}.  Second, even though the self energy may
be written in the form \be \Sigma_1={\rm Tr}{\cal T}_1G_3^0,
\label{SigmaG} \ee with the $T$-matrix calculated to second order in
the ladder approximation, it leads to an expression of the form
(\ref{sigma}) where the $T$-matrix contains both ladder and bubble
contributions.    Only for  nondegenerate 1-fermions are the bubble
diagrams unimportant compared with the ladder diagrams because
$f^1\ll1$. By extending the above arguments to higher-order terms,
one sees that  the self energy calculated from Eq.\ (\ref{SigmaG})
with ${\cal T}_1$ calculated in the ladder approximation agrees with
Eq.\ (\ref{sigma}) only for non-degenerate 1-fermions.

\subsection{Arbitrary order}
In higher-order processes, a qualitatively new feature appears: to
obtain the result (\ref{sigma}), with the $T$-matrix given by Eq.\
(\ref{lippmann}), one cannot use Eq.\ (\ref{SigmaG}) with the
$T$-matrix calculated in the ladder approximation. To obtain the
correct result for $m/m_3\rightarrow 0$, the $n$th order
contribution to the self energy has the form
\begin{gather}
\Sigma_{1}^{(n)}(\tau_b-\tau_a)=
-(-V_{1})^n\int_0^\beta d\tau_1 \ldots d\tau_{n-2}\times\nonumber\\
\hspace{-4em} G_{1}(\tau_b-\tau_{n-2})G_{1}(\tau_{n-2}-\tau_{n-3})\ldots G_{1}(\tau_1-\tau_a)\times\nonumber\\
[ G_{3}(\tau_b-\tau_{n-2})G_{3}(\tau_{n-2}-\tau_{n-3})\ldots G_{3}(\tau_1-\tau_a)G_{3}(\tau_a-\tau_b)\nonumber\\
+\text{ all } \tau \text{ permutations}].
\label{nth}
 \end{gather}
The momentum  sums are suppressed for the moment to highlight the essential parts of the reasoning.
We refer to this term as an $n$th order contribution, even though we use renormalized propagators, which may give rise to contributions of higher order in $V_1$ when expressed in terms of bare propagators.  The first term in the square brackets  corresponds to Eq.\ (\ref{SigmaG}) with $\cal T$ calculated is the ladder approximation, and it
does not give the result (\ref{sigma}) for the self energy in the limit $m_3/m\gg1$.  However, if one adds to it contributions corresponding to all possible ways of attaching the
interactions occurring at times $\tau_1\ldots\tau_{n-2}$ to the 3-bubble  with vertices at  $\tau_a$ and $\tau_b$, one does indeed
recover the result  (\ref{sigma}). This procedure is indicated by the last line of (\ref{nth}) and
 yields in third order the two diagrams depicted in Fig.\ \ref{Sigmadiagrams}(b).  The structure of a
  typical high order diagram generated in this way
  is illustrated in Fig.\ \ref{Sigmadiagrams}(c). To show that (\ref{nth}) produces the correct impurity result,  we use the fact that
  the non-interacting 3-propagator for  $m/m_3\rightarrow 0$ becomes
  \begin{equation}
G^0_{3}(p,\tau)=\left\{\begin{array}{lc}
 f_p^3 e^{-(\epsilon_3 - \mu_3)\tau} &\text{ for } \tau<0\\
-e^{-(\epsilon_3 - \mu_3)\tau}  &\text{ for } \tau>0,
\end{array}
\right.
\label{infmass}
\end{equation}
where $\tau<0$ corresponds to the propagation of a 3-hole and  $\tau>0$ to the propagation of a 3-particle.  Self energy contributions to $G_3$ can be absorbed
in the chemical  potential $\mu_3$ for low $T$ and $m/m_3\ll 1$.  The point is that for any value of  $\tau_a,\tau_b, \tau_1\ldots\tau_{n-2}$ between $0$ and $\beta$,
only one term inside the square brackets in (\ref{nth}) will have one hole propagator  and   $n-1$ propagators for the 3-particles. This term will scale as $\sim f^3_q$.
 All other terms have at least two hole propagators and will be suppressed in the limit of low concentration of the 3-particles.
 Note that the ladder diagram is not enough to include the leading order diagram in $f^3$
 for any value of  $\tau_a,\tau_b, \tau_1\ldots\tau_{n-2}$: one has to include diagrams
corresponding to all possible ways of attaching   $\tau_1\ldots\tau_{n-2}$ to the 3-bubble.
 In the limit $m/m_3\rightarrow0$,  the momentum integrals in (\ref{nth}) decouple
and the integral over the 3-hole line yields the density $n_3$. We obtain
\begin{gather}
\Sigma_{1}^{(n)}(\tau_b-\tau_a)=
n_3V_{1}^n\int_0^\beta d\tau_2 \ldots d\tau_{n-1}\times\nonumber\\
G_{1}(\tau_b-\tau_{n-2})G_{1}(\tau_{n-2}-\tau_{n-3})\ldots G_{1}(\tau_1-\tau_a).
\label{minfnth}
\end{gather}
In frequency space this reads
\begin{equation}
 \Sigma_{1}^{(n)}(\omega)=
 n_3V_{1} \left[V_{1}G_{1}(\omega)\right]^{n-1}.
 \end{equation}
 Summing all orders for $\Sigma$ gives
 \begin{equation}
 \Sigma_{1}(\omega)=n_3[1-V_{1}G_{1}(\omega)]^{-1}V_1=n_3 {\mathcal{T}}_{1}(\omega).
 \end{equation}
 This agrees with (\ref{sigma}) and we have shown that one recovers the correct impurity result for the self energy  in the limit of $n_3$ small and
  $m/m_3\rightarrow 0$, when all crossed diagrams of the
type   illustrated in Fig.\ \ref{Sigmadiagrams}(c) are included.

The same argument applies to vertex corrections. Consider $1$ and $2$ fermions  simultaneously scattering on a 3-particle.
A typical diagram needed to be included to recover the correct impurity result is shown in Fig.\ \ref{Sigmadiagrams}(d):
For an $n$th order diagram, one has to include all possible ways of attaching the interactions occurring at   $\tau_1\ldots \tau_{n}$ to the $n$ propagators in the 3-loop.
In this way, the term where there is only one hole in  the 3-loop  is included for any value of the time arguments. This term scales as
 $n_3$ whereas all other diagrams are suppressed by higher powers of $n_3$. When these diagrams are included to all orders, the
effective interaction between a 1-fermion and a 2-fermion both scattering on a 3-atom becomes
\begin{gather}
V_{\rm eff}=n_3[1-V_2 G_{2}(\omega_2)]^{-1}V_{2}
[1-V_1G_1(\omega_1)]^{-1}V_1\nonumber \\ =
n_3 {\mathcal{T}}_{1}(\omega_1) {\mathcal{T}}_{2}(\omega_2).
\end{gather}
This agrees with the impurity scattering result given by (\ref{veff}).

  \subsection{Higher loops and the X-ray edge problem}

  So far we have considered diagrams in which there is a single fermion loop containing fermions in states 1 and 2.  We now comment on the effect of including contributions with a higher number of loops.  The problem under consideration in this paper has a number of points in common with the X-ray edge problem, where the contributions from terms containing many fermion loops change qualitatively the nature of the threshold behavior \cite{mahan, NDD} from a step function at the Fermi surface when a single fermion loop is included to a power law whose exponent depends on the phase shift for scattering of an electron in the conduction band from a deep hole.  In the X-ray edge problem, conduction electrons scatter from a deep hole, which is present only for times between that at which the electron--deep-hole pair is
  created and that at which it is destroyed.
 The complications in the X-ray edge problem are due to the fact that the higher-order loop contributions depend
 on the times at which the particle-hole pair is created and destroyed.
 In the problem under consideration in this paper, however, the heavy atoms in the state 3 are present for all time.   The effect of the higher-order loops is simply to renormalize the propagator for a 3-atom.  The self energy of a 3-atom depends on energy but, within the approximation of a short-range potential made above,  is  independent of momentum.  When higher loop contributions are included,
 the chemical potential of the impurities must be adjusted so that the number of impurities is equal to the required value.

  \subsection{Analogy with ``shadowing'' in nuclear physics}

  The result (\ref{Lorentz}) has a simple interpretation, since it is equivalent to the statement that the self-energy of a particle-hole pair due to interaction with an impurity is proportional to $e^{2i(\delta_2-\delta_1)}-1$.  Since the self energy is proportional to the $T$-matrix for scattering of a pair from an impurity, which is in turn  proportional to $S-1$, where $S$ is the corresponding $S$-matrix, this implies that  $S=e^{2i(\delta_2-\delta_1)}$.  In physical terms, this says that the extra phase acquired by the pair is the sum of the phase changes experienced by a particle in state 2 and a hole in state 1.    The reason that vertex corrections, which correspond to interference terms, are so important in the present problem is that the external field creates a particle and the hole at the same point in space.  Thus if, say, the particle is close to an impurity, the hole will also be close to an impurity.   If only self-energy corrections are included, this is equivalent to assuming that the particle and the hole are uncorrelated in space.

 Insight into the result for the line shift may be obtained by making use of the identity
 \vspace{-2em}
 \be
 \sin 2(\delta_1-\delta_2)=\sin 2\delta_1(1-2\sin^2\delta_2)-\sin 2\delta_2(1-2\sin^2\delta_1),
 \ee
 which implies that the energy shift is given by the real parts of the self energy of a 1-fermion and a 2-hole, multiplied by factors $1-2\sin^2\delta$.  To interpret this result, we observe that
  the total cross section for scattering of a 3-atom by a $\sigma$-fermion is proportional to  $\sin^2\delta_\sigma/k^2$, where $k$ is the wavenumber of the atom.  If one changes ones perspective and regards the process as the interaction of an impurity fermion with a particle-hole pair, this equation implies that the amplitude of  an impurity fermion at the position of the 2-hole is reduced by an amount $\sim \sin^2\delta_1$ due to scattering from the 1-fermion, and likewise for the amplitude of the impurity at the 2-hole.
This is reminiscent of the experimental observation that the total cross section for scattering of pions from   deuterons is less than the sum of the cross sections for scattering of a pion from a single neutron and a single proton, a phenomena referred to as ``shadowing''.   It reflects the fact that the neutron and proton in the deuteron are correlated, and therefore the pion field incident on, e.g., the proton is reduced by scattering from the neutron \cite{glauber}.  The analogy between the two situations is not complete, however, since in the problem considered by Glauber the wavelength of the pion is small compared with the separation of the neutron and proton in the deuteron, while in the problem under investigation here
the wavelength of the particle and the hole is large compared with their separation, which is initially zero.  As a consequence, where $k^2$ appears in the present problem, this is replaced by a factor $\propto <1/r^2>$, the average of the inverse square of the separation of the neutron and proton in the deuteron.

 \section{Concluding remarks}

 In this paper we have solved a simple model for clock shifts for hyperfine transitions between states of a fermionic atom, in the presence of a low density of much more massive atoms.  The calculation shows the importance of vertex corrections, which completely change the dependence of the shift and the width of the clock transition on the scattering phase shifts.    The calculations are valid for bystander atoms, either fermionic or bosonic, which are very much more massive than the majority fermions, and an important problem for the future is to study a finite mass ratio.

Throughout, we have have neglected the interaction between 1-atoms and 2-atoms.  When there are no bystanders, the transition has no shift and no width, and this result also holds in the presence of bystanders, provided the 1-3 and 2-3 interactions are identical, since in that case the SU(2) invariance still holds.    However, when the 1-3 and 2-3 interactions are different, the line shift and width can be affected by the 1-2 interaction, which is an unexplored effect in the cold atomic gas context.

The calculations indicate that experiments on clock shifts in mixtures of atoms with different masses would be useful.
Since pairing correlations are suppressed when species have very different concentrations, these would enable one to obtain information about correlations in a state less complicated than a paired superfluid.

 An important theoretical result of our calculations is that it is generally not sufficient to include just ladder diagrams, since in the case considered, particle-hole correlations are necessary in order to recover the correct result for a low density of the minority component.

 \acknowledgements

 We are grateful to H.\ T.\ C.\ Stoof for seminal discussions at ECT$^*$ in Trento (Italy), and GMB and CJP would like to thank ECT* for hospitality. We thank Wolfgang G\"otze for helpful correspondence.


\begin{thebibliography}{20}

\bibitem{balling} L.\ C.\ Balling, R.\ J.\ Hanson, and F.\ M.\ Pipkin, Phys.\ Rev.\ {\bf 133}, A607 (1964); Erratum, ibid {\bf 135}, AB1 (1964).

\bibitem{koelman}    J.\ M.\ V.\ A.\ Koelman, S.\ B.\ Crampton, H.\ T.\ C.\ Stoof, O.\ J.\ Luiten,
       and B.\ J.\ Verhaar,
Phys. Rev. A {\bf 38}, 3535 (1988).

 \bibitem{gupta} S.\ Gupta, Z.\ Hadzibabic, M.\ W.\  Zwierlein, C.\ A.\   Stan, K.\
Dieckmann, C.\ H.\  Schunck, E.\ G.\ M.\ van Kempen, B.\ J.\  Verhaar, and W.\ Ketterle,
Science \textbf{300}, 1723 (2003).

  \bibitem{mit_two} M.\ W.\ Zwierlein, Z.\ Hadzibabic, S.\ Gupta and W.\
Ketterle, Phys.\ Rev.\ Lett.\ \textbf{91}, 250404 (2003).

    \bibitem{chin_rf} C.\ Chin, M.\  Bartenstein, A.\  Altmeyer, S.\  Riedl, S.\
Jochim, J.\  H.\  Denschlag, and R.\  Grimm, Science \textbf{305}, 1128 (2004).


\bibitem{shin} Y.\ Shin, C.\ H.\  Schunck, A.\ Schirotzek, and W.\ Ketterle,  Phys. Rev. Lett. {\textbf 99}, 090403 (2007).
C.\ H.\ Schunck, Y.\  Shin, A.\  Schirotzek, and W.\ Ketterle,  Nature {\textbf 454}, 739 (2008).

\bibitem{kinnunen} J.\ Kinnunen, M.\ Rodriguez, and P.\  T\"orm\"a,
Science   \textbf{305}, 1131  (2004).

\bibitem{yu} Z.\ Yu and G.\ Baym, Phys.\ Rev.\ A {\bf 73}, 063601 (2006).

\bibitem{perali} A.\ Perali, P.\  Pieri,  and G.\ C.\ Strinati, Phys. Rev. Lett. \textbf{100}, 010402 (2008).

\bibitem{pieri}  P.\  Pieri, A.\ Perali, and G.\ C.\ Strinati, Nature Physics {\bf 5}, 736 (2009).

\bibitem{levin} Y.\ He, C.-C. Chien, Q.\  Chen, and K.\ Levin,  Phys. Rev. Lett. \textbf{102}, 020402 (2009).

\bibitem{baymzwierlein}
G.\ Baym, C.\ J.\ Pethick, Z.\ Yu, and M.\ W.\ Zwierlein, Phys.\ Rev.\ Lett.\ {\bf 99}, 190407 (2007).

\bibitem{zwerger} M.\ Punk and W.\ Zwerger,
Phys.\ Rev.\ Lett.\ {\bf  99}, 170404 (2007).

\bibitem{bruun1} P.\ Massignan, G.\ M.\ Bruun, and H.\ T.\ C.\ Stoof,
Phys. Rev. A {\bf 77}, 031601 (2008).

\bibitem{bruun2} P.\ Massignan, G.\ M.\ Bruun, and H.\ T.\ C.\ Stoof
Phys. Rev. A {\bf 78}, 031602 (2008).

\bibitem{mueller} S.\ Basu and E.\ J.\ Mueller, Phys.\ Rev.\ Lett.\ {\bf 101}, 060405 (2008).

\bibitem{AL} L.\ G.\ Aslamazov and A.\ I.\ Larkin, Phys.\ Lett.\ \textbf{26A}, 228 (1968).

\bibitem{glauber} R. J. Glauber, Phys. Rev. {\bf 100}, 242 (1955).

\bibitem{AGD} A.\ A.\ Abrikosov, L.\ P.\ Gorkov, and I. Ye.\ Dzyaloshinskii, {\it Quantum Field Theoretical Methods in Statistical Physics}, (Pergamon, Oxford, 1965), \S 39.

\bibitem{baym}  G. Baym, Phys.\ Rev.\ {\bf 127}, 1391 (1962).

\bibitem{mahan} G.\ D.\ Mahan, Phys.\ Rev.\   {\bf 163}, 612  (1967).

\bibitem{NDD} P. Nozi{\`e}res and C. T. De Dominicis, Phys.\ Rev.\   {\bf 178}, 612  (1969).

 \bibitem{astr} G.\ E.\ Astrakharchik, J.\ Boronat, J.\ Casulleras, and S.\ Giorgini, Phys.\ Rev.\ Lett.\ \textbf{93}, 200404 (2004).

\bibitem{bruunbaym2} C. Chin and P. Julienne, Phys. Rev. A \textbf {71}, 012713 (2005); G. M. Bruun and G. Baym, Phys.\ Rev.\ A \textbf{74}, 033623 (2006).


\bibitem{bartenstein} M.\ Bartenstein, A.\  Altmeyer, S.\ Riedl, R.\ Geursen, S.\ Jochim, C.\ Chin, J.\ H.\ Denschlag, R.\ Grimm, A.\ Simoni, E.\ Tiesinga, C.\ J.\ Williams,  and P.\ S.\ Julienne,  Phys. Rev. Lett. {\bf 94}, 103201 (2005).


\bibitem{KB}L.\ P.\ Kadanoff and G.\  Baym,
{\it Quantum Statistical Mechanics}, (Benjamin, New York, 1962).

\bibitem{Keldysh} L.\  V.\  Keldysh, Zh.\ Eksp.\ Teor.\ Fiz.\ {\bf 47}, 1515 (1964) --JETP {\bf 20}, 1018 (1965).



\end{thebibliography}
\end{document}